\documentclass[twocolumn,twoside,leqno]{article}
\usepackage{preamble}

\title{\blockFIFO{} \& MultiFIFO:\@ Scalable Relaxed Queues}

 \author{Stefan Koch\thanks{Karlsruhe Institut of Technology, Germany (\email{stefan.koch@student.kit.edu}, \email{sanders@kit.edu}, \email{williams@kit.edu}).}
   \and Peter Sanders\footnotemark[1]
 \and Marvin Williams\footnotemark[1]}

\begin{document}
\maketitle
\fancyfoot[R]{\scriptsize{Copyright \textcopyright\ 2026 by SIAM\\
Unauthorized reproduction of this article is prohibited}}
\begin{abstract}
    FIFO queues are a fundamental data structure used in a wide range of applications.
    Concurrent FIFO queues allow multiple execution threads to access the queue simultaneously.
    Maintaining strict FIFO semantics in concurrent queues leads to low throughput due to high contention at the head and tail of the queue.
    By relaxing the FIFO semantics to allow some reordering of elements, it becomes possible to achieve much higher scalability.
    This work presents two orthogonal designs for relaxed concurrent FIFO queues, one derived from the MultiQueue and the other based on ring buffers.
    We evaluate both designs extensively on various micro-benchmarks and a breadth-first search application on large graphs.
    Both designs outperform state-of-the-art relaxed and strict FIFO queues, achieving higher throughput and better scalability.
\end{abstract}
\section{Introduction.}
FIFO (First-in first-out) queues are data structures that support the insertion of elements (\Op{push}) and the deletion of the least recently inserted element (\Op{pop}).
They are at the core of a wide range of applications, including breadth-first search, processing pipelines, message queues, and network routers.
Sequential implementations using circular arrays are fast and cache-efficient, offering high throughput with constant-time operations.
As most performance-critical computations today use parallel hardware, concurrent queues have become increasingly important.
Indeed, in many applications, a concurrent queue serves as the central coordination mechanism among threads.
For example, multiple threads may concurrently produce and consume data items from the queue, or, in a breadth-first traversal of a graph, the queue may hold the nodes yet to be explored.

Ideally, using $p$ threads, one would like the throughput to be close to $p$ times the throughput of a sequential queue.
Unfortunately, high contention on the head and tail pointers of the queue is unavoidable~\cite{ellenInherentSequentialityConcurrent2012, attiyaLawsOrderExpensive2011}, and even sophisticated implementations of concurrent queues with strict semantics (e.g.,~\cite{michaelSimpleFastPractical1996,morrisonFastConcurrentQueues2013,romanovStateoftheArtLCRQConcurrent2023}) suffer from poor scalability.
The throughput remains constant or might even deteriorate with increasing numbers of threads.
Thus, such designs are not suitable for situations that require a high throughput with many threads.

In order to achieve higher scalability, the idea of relaxing the semantics of queues and similar data structures in order to benefit from reduced contention has become an increasingly relevant topic of study~\cite{afekQuasiLinearizabilityRelaxedConsistency2010, kirschFastScalableLockFree2013, henzingerQuantitativeRelaxationConcurrent2013, vongeijerBalancedAllocationsEfficient2025, williamsEngineeringMultiQueuesFast2025}.
We say that a popped element has \emph{rank error} $r$ when it was pushed later than $r$ other elements that are still in the queue.

In this paper, we introduce the \emph{MultiFIFO} and the \emph{\blockFIFO{}}, two concurrent relaxed FIFO queues with particularly good scalability.
The MultiFIFO (see \cref{s:multiFIFO}) is an adaptation of the MultiQueue~\cite{rihaniMultiQueuesSimpleRelaxed2015,williamsEngineeringMultiQueuesFast2021,williamsEngineeringMultiQueuesFast2025}, a state-of-the-art relaxed priority queue that builds upon multiple internal strict queues.
Pushed elements go to a random queue, along with their insertion time stamp.
Pops remove the least recent element from two randomly chosen queues.
The MultiFIFO has expected constant time per operation and inherits a rank error linear in the number of threads from the MultiQueue.
By making threads reuse the same internal queues for a fixed number of consecutive operations, performance can be further increased at the cost of higher rank errors.

The \blockFIFO{} (see \cref{s:blockFIFO}) is based on a ring buffer of small FIFO \emph{blocks}, where threads operate on distinct blocks to reduce contention.
It exhibits even higher throughput than the MultiFIFO in its most relaxed configurations.

\paragraph*{Summary of Contributions.}
\begin{itemize}
    \item Introduction of two highly scalable relaxed concurrent FIFOs: the MultiFIFO and the \blockFIFO{}.
    \item Extensive experiments (\cref{s:eval}) on different architectures with various benchmarks indicating an order-of-magnitude improved throughput of the new FIFO queues compared to the previous state of the art.
\end{itemize}

\section{Preliminaries.}
The first-in first-out (FIFO) queue is a data structure that manages a dynamic sequence of elements, supporting the following two operations:
\begin{description}
    \item[\Op{push}] Insert an element at the end (\emph{tail}) of the sequence.
    \item[\Op{pop}] Remove and return the element at the front (\emph{head}) of the sequence.
\end{description}
If a FIFO is empty, the \Op{pop} operation \emph{fails} and returns the special element $\bot$, which cannot be inserted.
\emph{Bounded} FIFO queues have a fixed capacity of elements, and pushing into to a full bounded FIFO conventionally fails\footnote{Other semantics, such as overwriting the oldest element or allocating more space, are possible.}.
The \emph{ring buffer} (or \emph{circular buffer}) is a common bounded FIFO queue, which conceptually arranges the slots for elements in a circular manner.
Ring buffers are typically implemented as contiguous fixed-size arrays with head and tail pointers indicating the next slot to pop from and push into, respectively.
These pointers are incremented by the corresponding operations and wrap around to the beginning when they reach the end of the array.
\emph{Concurrent} FIFO queues allow concurrent \Op{push} and \Op{pop} operations by multiple threads.

\emph{Linearizability}~\cite{herlihyLinearizabilityCorrectnessCondition1990} is a popular consistency criterion for specifying the semantics of concurrent data structures with respect to their sequential counterparts.
Essentially, a concurrent data structure is linearizable if each operation appears to execute instantly at a single point in time (its \emph{linearization point}) between its invocation and response.
Linearizability is generally highly desirable: with non-overlapping operations, the data structure behaves like a sequential one, and concurrent executions offer strong and intuitive guarantees.
Unfortunately, linearizable FIFO queue implementations suffer from high contention and exhibit limited scalability~\cite{ellenInherentSequentialityConcurrent2012, attiyaLawsOrderExpensive2011}.
To alleviate this issue, the semantics of FIFO queues can be \emph{relaxed}, allowing the \Op{pop} operation to remove elements out-of-order, or even \emph{fail} to remove elements that are still in the queue.
For many applications, it is useful for failed pops to be linearizable, meaning that popping may only fail if the queue was empty at some point during the operation.
A natural quality metric for the degree of relaxation is the \emph{rank error}: For an element $e$ that was returned by a \Op{pop} operation, the rank error of the \Op{pop} operation is the number of elements in the queue that were inserted before $e$.

An algorithm is \emph{lock-free} if at least one thread is guaranteed to make progress towards completing an operation within a bounded number of steps, regardless of the actions of other threads.

The \emph{ABA problem} arises when a thread reads the same value $A$ from a shared memory location twice, incorrectly assuming the value has not changed between the reads, even though it was changed to $B$ and then back to $A$ by other threads.

We denote $p$ as the (maximum) number of threads participating in an algorithm concurrently.
An event occurs \emph{with high probability} (with respect to $p$) if the probability is at least $1-p^{-a}$ for some constant $a \geq 1$.

\section{Related Work.}
The FIFO queue is a conceptually simple data structure that appears in most introductory algorithm textbooks (e.g.,~\cite{cormenIntroductionAlgorithmsFourth2022}) and is readily available in the standard library of most programming languages.
As a core component in a wide range of applications, including message queues, processing pipelines, and breadth-first searches, FIFO queues have been studied extensively in the literature.
Here, we focus on concurrent FIFO queues with an emphasis on relaxed variants.
\subsection*{Linearizable FIFO Queues.}
One of the earliest linearizable lock-free concurrent FIFO queue algorithms is the \emph{MS-Queue}, proposed by Michael and Scott~\cite{michaelSimpleFastPractical1996}.
The algorithm is based on a linked list, where nodes can be appended and removed in a lock-free manner.
To avoid the ABA problem during atomic updates of next-pointers, each pointer is tagged with a version number.
The design suffers from high contention on the head and tail pointers, which limits its scalability.
Nevertheless, its simplicity and ease of implementation have made it the foundation for numerous subsequent designs (e.g.,~\cite{haasDistributedQueuesShared2013,gidenstamCacheAwareLockFreeQueues2010,hoffmanBasketsQueue2007}).
One notable example is the \emph{Baskets Queue} by Hoffman et~al\@.~\cite{hoffmanBasketsQueue2007}, which uses unordered \emph{baskets} for concurrently inserted elements as nodes in the linked list.

Tsigas and Zhang~\cite{tsigasSimpleFastScalable2001} propose a bounded, lock-free FIFO queue based on ring buffers.
Although bounded queues are less flexible than unbounded ones, they offer several practical advantages: they do not require dynamic memory allocation, are generally more cache-friendly, and do not require complex memory reclamation mechanisms.

The LCRQ (List of Concurrent Ring Queues), introduced by Morrison and Afek~\cite{morrisonFastConcurrentQueues2013}, is a state-of-the-art concurrent FIFO queue design.
It combines the MS-Queue with ring buffers to leverage the advantages of both data structures.
Its implementation favors the more efficient atomic fetch-and-add operations over compare-and-swap operations.
Both the MS-Queue and the LCRQ rely on the double-width compare-and-swap operation (dCAS), which most platforms do not support natively.
The LPRQ (\emph{portable} LCRQ)~\cite{romanovStateoftheArtLCRQConcurrent2023} improves the portability of the LCRQ by eliminating the dCAS while maintaining comparable performance.

\subsection*{Relaxed FIFO Queues.}
Afek et~al\@.~\cite{afekQuasiLinearizabilityRelaxedConsistency2010} introduce the \emph{Segmented Queue}, a relaxed FIFO queue based on a linked list of \emph{segments}, where each segment is a static array of size $C$.
When pushing, a thread writes the element into a random empty cell in the tail segment, appending a new segment if necessary; popping is performed analogously.
The Segmented Queue exhibits bounded worst-case rank errors in $\BigO(C)$.
The \emph{k-FIFO} queue by Kirsch et~al\@.~\cite{kirschFastScalableLockFree2013} enhance this design with scalable, linearizable emptiness checks.

A popular approach to relaxing the semantics of a data structure is to employ multiple (thread-safe) instances of the data structure and distribute the access to them among the threads (e.g.,~\cite{haasDistributedQueuesShared2013,rukundoMonotonicallyRelaxingConcurrent2019,vongeijerBalancedAllocationsEfficient2025}).
The distribution mechanism is crucial for the performance and quality guarantees of these designs.
Rukundo et~al\@.~\cite{rukundoMonotonicallyRelaxingConcurrent2019} introduce the \emph{2D-Queue}, which distributes operations among its internal queues such that their sizes remain within a fixed range.
This design guarantees a bounded worst-case rank error of $\BigO(wr)$, where $w$ is the number of queue instances and $r$ is the length of the range.
The $d$-RA ($d$-randomized load balancer) by Haas et~al\@.~\cite{haasDistributedQueuesShared2013} samples $d\geq 1$ data structure instances for each operation, selecting the least-loaded for pushes and the most-loaded for pops.
However, von Geijer et~al\@.~\cite{vongeijerBalancedAllocationsEfficient2025} demonstrate that this balancing scheme can lead to increasing rank errors with growing queue sizes.
They propose the \emph{$d$-CBO} ($d$-Choice Balanced Operations), which balances using dedicated push and pop counters per queue instance.
A \Op{push} operation samples $d\geq 2$ instances and chooses the one with the fewest prior pushes.
Symmetrically, a \Op{pop} operation samples $d$ instances and chooses the one with the fewest prior pops.
Interestingly, this scheme stabilizes the rank errors empirically. Concurrent bags~\cite{sundellLockfreeAlgorithmConcurrent2011} impose no ordering on the elements, thus can be viewed as the most extreme version of relaxed FIFO queues.

Henzinger et~al\@.~\cite{henzingerQuantitativeRelaxationConcurrent2013} introduce the \emph{Quantitative Relaxation} framework to specify and analyze semantics of relaxed concurrent data structures.
However, the framework is not applicable to randomized algorithms where the rank errors are not bounded.

\section{The MultiFIFO.}\label{s:multiFIFO}
The MultiQueue~\cite{rihaniMultiQueuesSimpleRelaxed2015,williamsEngineeringMultiQueuesFast2021,williamsEngineeringMultiQueuesFast2025} is a state-of-the-art relaxed concurrent priority queue.

It consists of an array of $c\cdot p$ sequential priority queues, each protected by an mutual exclusion lock, where $c\geq2$ is the \emph{queue factor}.
The operations are handled asymmetrically:
An insertion randomly selects and locks a single queue to insert the new element.
A deletion uses the \emph{power of two choices} principle by randomly sampling \emph{two} queues, and locking the one containing the element with the highest priority to perform the deletion.
If a thread fails to acquire a lock during either operation, it retries the operation.

Despite employing locks, the MultiQueue is \emph{probabilistically wait-free}, meaning that all operations of all threads make progress within a bounded number of steps in expectation.
To improve the performance of the MultiQueue, Williams et~al\@.~\cite{williamsEngineeringMultiQueuesFast2021} propose the concept of \emph{stickiness}, where threads reuse the same two queues for $s$ (the \emph{stickiness period}) consecutive operations.
For details on the MultiQueue, we refer to the original paper~\cite{williamsEngineeringMultiQueuesFast2021,williamsEngineeringMultiQueuesFast2025}.

A promising approach to creating a relaxed FIFO queue is to adapt the MultiQueue.
For this adaptation, which we call the \emph{MultiFIFO}, we replace the sequential priority queues with in-place ring buffers and tag each element with its insertion timestamp.
The pop operation randomly samples two ring buffers, compares their head elements, and removes the one with the earlier timestamp.

Due to the similarity to the MultiQueue, the MultiFIFO inherits many of its desirable properties, such as \emph{probabilistic wait-freedom}~\cite{williamsEngineeringMultiQueuesFast2025}, linear (in $p$) expected rank errors of $\tfrac{5}{6}cp-1+\tfrac{1}{6cp}$~(see \cite{walzerSimpleExactAnalysis2025}), and rank errors in $\BigO(p\log p)$ with high probability.
Both operations are in $\BigO(1)$ in expectation, since finding an unlocked ring buffer takes constant time in expectation (see~\cite{williamsEngineeringMultiQueuesFast2025}) and the time for the actual operations on the ring buffers is constant.

\section{The \blockFIFO{}.}\label{s:blockFIFO}
In this section, we present the \blockFIFO{}, a bounded, lock-free, relaxed concurrent FIFO queue.
We begin by outlining its basic design, then analyze its practical limitations and present several key improvements.

\subsection{Basic Design.}
\begin{figure}[H]
    \centering
    \includegraphics{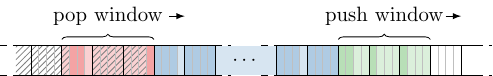}
    \caption{%
        Schematic diagram of the \blockFIFO{} with block size $C=4$ and window size $w=3$, represented as a linear array.
        The colored area represents the ``active'' part of the array.
        Slots with stronger colors contain elements.
        The line pattern indicates that a slot is emptied and not used again.%
    }%
    \label{fig:blockFIFO}
\end{figure}

Conceptually, the \blockFIFO{} is an infinite array of \emph{blocks}, where each block is a small, fixed-size buffer of size $C$, which holds the actual elements.
For the sake of simplicity, we present the data structure assuming an infinite, linear memory layout.
We show a practical, lock-free implementation based on ring buffers in \cref{s:bf-impl}.

A block can be in one of three states: \emph{unclaimed}, \emph{claimed}, or \emph{closed}.
An unclaimed block is empty and can be claimed for a \Op{push} operation by any thread, after which it becomes claimed.
No other thread may claim a claimed block or insert elements into it.
Once all elements have been popped from a block, it is closed and can no longer be used.
The \emph{push window} and the \emph{pop window} are the ranges of blocks that are currently available for \Op{push} and \Op{pop} operations, respectively.
The \emph{window size} $w$ (the number of blocks in each window) is linear in the number of threads $p$, resulting in $w=B\cdot p$ blocks, where $B\geq 1$ is the \emph{block factor}.
Initially, all blocks are unclaimed and the pop window is positioned directly behind the push window.
Figure~\ref{fig:blockFIFO} shows the \blockFIFO{} with $C=4$ and $w=3$ schematically.

\begin{algorithm2e}[t]
    \caption{Pseudocode for the \Op{push} and \Op{pop} operations.
        The \Op{pop} operation returns the deleted element or $\bot$ if the queue was empty.}\label{alg:ops_linear}
    \SetKwFunction{Push}{push}
    \SetKwFunction{InsertInBlock}{pushToBlock}
    \SetKwFunction{ClaimBlock}{claimNewBlock}
    \SetKwFunction{AdvancePushWindow}{advancePushWindow}
    \SetKwFunction{CurrentPushWindow}{currentPushWindow}
    \SetKwFunction{ClaimedBlock}{lastClaimedBlock}
    \SetKwData{Win}{$z$}
    \SetKwData{Elem}{$e$}
    \SetKwData{Block}{$k$}
    \SetKwFunction{Pop}{pop}
    \SetKwFunction{DeleteFromBlock}{popFromBlock}
    \SetKwFunction{FindUnclosedBlock}{findPopBlock}
    \SetKwFunction{AdvancePopWindow}{advancePopWindow}
    \SetKwFunction{CurrentPopWindow}{currentPopWindow}
    \SetKwFunction{LastPopBlock}{lastPopBlock}
    \SetKwFunction{IsEmpty}{isEmpty}
    \SetKwData{Block}{$k$}
    \SetKwData{Elem}{$e$}
    \SetKwData{Win}{$z$}
    \Fn{\Push{\Elem}}{
        $\Win \gets \CurrentPushWindow{}$\;
        $\Block \gets \ClaimedBlock{}$\;
        \While{$\neg\InsertInBlock{\Win, \Block, \Elem}$}{
            \While{$\neg(\Block \gets \ClaimBlock{\Win})$}{
                $\Win \gets \AdvancePushWindow{\Win}$\;
            }
        }
    }
    \;
    \Fn{\Pop{}}{
        $\Win \gets \CurrentPopWindow{}$\;
        $\Block \gets \LastPopBlock{}$\;
        \While{$(\Elem \gets \DeleteFromBlock{\Block}) = \Empty$}{
          \RepeatInf{
            $\Win \gets \AdvancePopWindow{}$\;
            \lIf{$\Block \gets \FindUnclosedBlock{\Win}$}{\Break}
                \lIf{\IsEmpty{}}{\Return{\Empty}}
            }
        }
        \Return{\Elem}\;
    }
\end{algorithm2e}

Algorithm \ref{alg:ops_linear} gives high-level pseudocode of the \Op{push} and \Op{pop} operations.
When pushing an element, a thread first tries to reuse the last block it claimed.
The method \texttt{pushToBlock($z,k,e$)} appends the element $e$ to block $k$, failing if $z$ is not the current push window or block $k$ is full.
If it fails, the thread attempts to claim a new block in the current push window (\texttt{claimNewBlock($z$)}).
To find an unclaimed block, it linearly scans the push window, starting from a random position.
If there is no unclaimed block, the thread advances the push window by $w$ blocks and retries the scan.
The method \texttt{advancePushWindow($z$)} only advances the push window if it is currently $z$, otherwise it returns the current push window.
Note that not all blocks have to be full for the push window to advance.

When popping, a thread first tries to pop from the same block it used for the last \Op{pop} operation.
If the block is not closed, the block might still contain elements and the pop window did not move past it.
The method \texttt{popFromBlock($k$)} removes the head element from block $k$, returning $\bot$ if the block was already empty.
If the block is empty afterwards, it closes the block.

To find a new block to pop from, the thread first advances the pop window (\texttt{advancePopWindow()}) past any leading closed blocks, while ensuring the window does not overlap the push window.
The thread then scans the current pop window linearly for a non-closed block, starting from a random block in the window $z$ (\texttt{findPopBlock($z$)}).
In case the pop window is empty and cannot advance, the thread checks whether the push window contains any elements.
If it does, it advances both windows by the full window width and retries the operation.
If the push window is also empty, the queue is determined to be empty and the \Op{pop} fails (\texttt{IsEmpty()}).
Unlike pushes, different threads must be allowed to pop from the same block.

\subsection{Theoretical Properties.}
\paragraph*{Failed pops are linearizable.}
For a \Op{pop} operation to fail, the pop window must be positioned directly behind the push window, all blocks in the pop window must be closed, and the push window must be empty.
After checking these conditions, the thread verifies that the push window did not move in the meantime.
Since no new elements can be inserted into closed blocks and no elements can be deleted from the push window, the queue must have been empty just after scanning the pop window and finding only closed blocks.

\paragraph*{Lockfreeness.}
Both the \Op{push} and \Op{pop} operations are lockfree.
We sketch an argument in \cref{s:bf-impl}.
\paragraph*{Asymptotic Considerations.}
We do not have a complete analysis of the \blockFIFO{} yet.
However, the design is based on two principles: By using windows with $\Omega(p)$ blocks, threads can mostly work on ``their'' local insertion and deletion buffer blocks.
By using blocks of size $\Omega(p)$, expensive operations like moving windows and searching for blocks can be amortized over $\Omega(p)$ fast, local, cache-efficient operations on local blocks.
This implies amortized constant time operations at the price of rank errors of size $\Omega(p^2)$; we view it as likely that this is also an upper bound.

This reasoning breaks down when the queue is almost empty.
In that case expensive global operations dominate.
The \blockFIFO{} is designed for situations where the queue is sufficiently full ($\Omega(p^2)$ elements) most of the time.

\subsection{Practical Improvements.}
Searching for a new block to operate on can be expensive, especially when only few suitable blocks are available.
We therefore augment the data structure with a bitset, where a bit indicates whether the corresponding block has been closed.
When a thread claims a new block for pushing, it sets the corresponding bit to one.
When a thread empties a block by popping from it, it sets the corresponding bit to zero.
The bitset improves the cache locality for the block search, since the individual blocks do not need to be inspected.
Further, multiple bits can be inspected simultaneously with SIMD (single instruction, multiple data) techniques.
\Cref{s:bitset-details} gives details about the implementation of the bitset.

Another performance bottleneck in our design is the contention when multiple threads pop elements from the same block.
To alleviate this, we employ a \emph{lookahead} window of $w$ blocks in front of the current pop window.
A popping thread first searches for \emph{fresh} blocks in the pop window, i.e., blocks that have not yet been popped from by other threads, analogously to the \Op{push} operation.
If no fresh block is available, it searches for one in the lookahead window, before considering non-closed blocks in the push window.
Since the pop window can only advance when the first block in it is closed, another promising approach is to bias the random block selection towards the front of the pop window.
While this may increase the contention on the first blocks, it facilitates faster advancement of the pop window to make new blocks available faster.
We did not investigate this approach beyond preliminary experiments.

\section{Evaluation.}\label{s:eval}
We evaluate and compare the \blockFIFO{} and MultiFIFO with state-of-the-art FIFO queues across multiple workloads and different hardware architectures.
This includes micro-benchmarks and a concurrent breadth-first search on various graphs.

\subsection{Methodology.}
To measure the maximum throughput, we design the following two micro-benchmarks that aim to mimic practical workloads.
\begin{itemize}
    \item \textbf{Push-Pop}: Each thread repeatedly performs alternating \Op{push} and \Op{pop} operations for 5 seconds, keeping the number of elements in the queue constant.
          We choose a large enough capacity for the bounded queues and pre-fill all queues with sufficiently many elements to ensure that the queues never run full or empty.
          We measure throughput as the number of \Op{push}--\Op{pop} iterations per second.
    \item \textbf{Producer-Consumer}: Given a fixed number of threads, some perform \Op{push} operations, while the other perform \Op{pop} operations.
          We use the same queue size and capacity as in the push-pop benchmark.
          Depending on the configuration, the number of elements may grow or shrink, in extreme cases the queues may even run full or empty.
          We measure the throughput as the minimum of \Op{push} and \Op{pop} operations performed per second.

\end{itemize}

The elements in the queue are 64-bit integers, since this data type is supported by all implementations.\footnote{
    The $k$-FIFO implementation reserves some higher-order bits for internal tagging, which are unavailable for data.
    This has no practical implications on our experiments.}
Each experiment is repeated five times, and we report the mean.
The standard deviation is shown in plots as errors bars where significant.

To measure the rank errors exhibited by a queue, we take timestamps for each insertion and deletion.
After the benchmark, we reconstruct a global order of operations, which we replay sequentially.
While this process is not perfectly accurate, we deem it sufficient for our purposes.

\begin{table*}[t]
    \centering
    \caption{Hardware details of all machines used in experiments.}\label{exp:machines}
    \begin{tabular}{c c c c c}
        \toprule
                       & CPU (ISA)               & Sockets/Cores/Threads & Max. Clock Freq. & L1d/L2/L3 Cache\footnote{L1d and L2 cache are given per core. L3 cache is given per socket.} \\
        \midrule
        \texttt{AMD}   & EPYC 9684X (x86-64)     & 1/96/192              & \qty{3.7}{\GHz}  & \qty{32}{\kibi\byte}/\qty{1}{\mebi\byte}/\qty{1152}{\mebi\byte}                              \\
        \texttt{ARM}   & Neoverse-N1 (ARMv8.2-A) & 1/80/80               & \qty{3.0}{\GHz}  & \qty{64}{\kibi\byte}/\qty{1}{\mebi\byte}/-                                                   \\
        \texttt{Intel} & Xeon Gold 6138 (x86-64) & 4/80/160              & \qty{3.7}{\GHz}  & \qty{32}{\kibi\byte}/\qty{1}{\mebi\byte}/\qty{27.5}{\mebi\byte}                              \\
        \bottomrule
    \end{tabular}
\end{table*}
The \blockFIFO{}, the MultiFIFO and all experiments are implemented in standard C\text{++}20.
Competitors are implemented in C and C\texttt{++}, some using non-standard extensions.
All code is compiled with GCC~14.2.0, using the flags \texttt{-O3 -DNDEBUG}.
We pin execution threads to hardware threads to increase stability and consistency across experiments.
We use three different machines (\texttt{AMD}, \texttt{ARM}, and \texttt{Intel}) for our benchmarks.
Details to the hardware of these machines can be found in \cref{exp:machines}.
All machines run Rocky Linux 9.5 with Linux kernel version 5.14.
Unless specified otherwise, we use machine \texttt{AMD}.
The source code for our implementation, including the data structures and the benchmarks, is available online\footnote{\url{https://zenodo.org/records/17293832}}.

\subsection{Competitors.}
We compare the \blockFIFO{} and MultiFIFO with the following implementations of state-of-the-art competitors found in the literature.
While the main focus is on relaxed FIFO queues, we also include strict FIFO queues for reference.
Relaxed FIFO queues are expected to outperform strict FIFO queues in terms of throughput and scalability at the cost of rank errors.
\begin{description}
    \item[BF] The BlockFIFO with configurable block factor $B$ and block size $C$.
    \item[MF] The MultiFIFO with configurable queue factor $c$ and stickiness period $s$.
    \item[$\mathbf{k}$-FIFO] Implementation\footnote{\url{https://github.com/cksystemsgroup/scal}} of the $k$-FIFO~\cite{kirschFastScalableLockFree2013, haas2015scal} with configurable segment size $k$.
          The authors suggest a segment size of $k=p$.
    \item[$\mathbf{d}$-CBO] Implementation\footnote{\url{https://github.com/dcs-chalmers/semantic-relaxation-dcbo}} of the $d$-CBO~\cite{vongeijerBalancedAllocationsEfficient2025} with configurable number of sub-queues per thread $c$.
    \item[LCRQ] Implementation\footnote{\url{https://zenodo.org/records/7337237}} of the LCRQ~\cite{morrisonFastConcurrentQueues2013}.
    \item[FAAAQueue] Implementation\footnote{\url{https://concurrencyfreaks.blogspot.com/2016/11/faaarrayqueue-mpmc-lock-free-queue-part.html}} of the FAAAQueue.
\end{description}

\subsection{Micro-Benchmarks.}
\begin{figure*}
    \centering
    \includegraphics{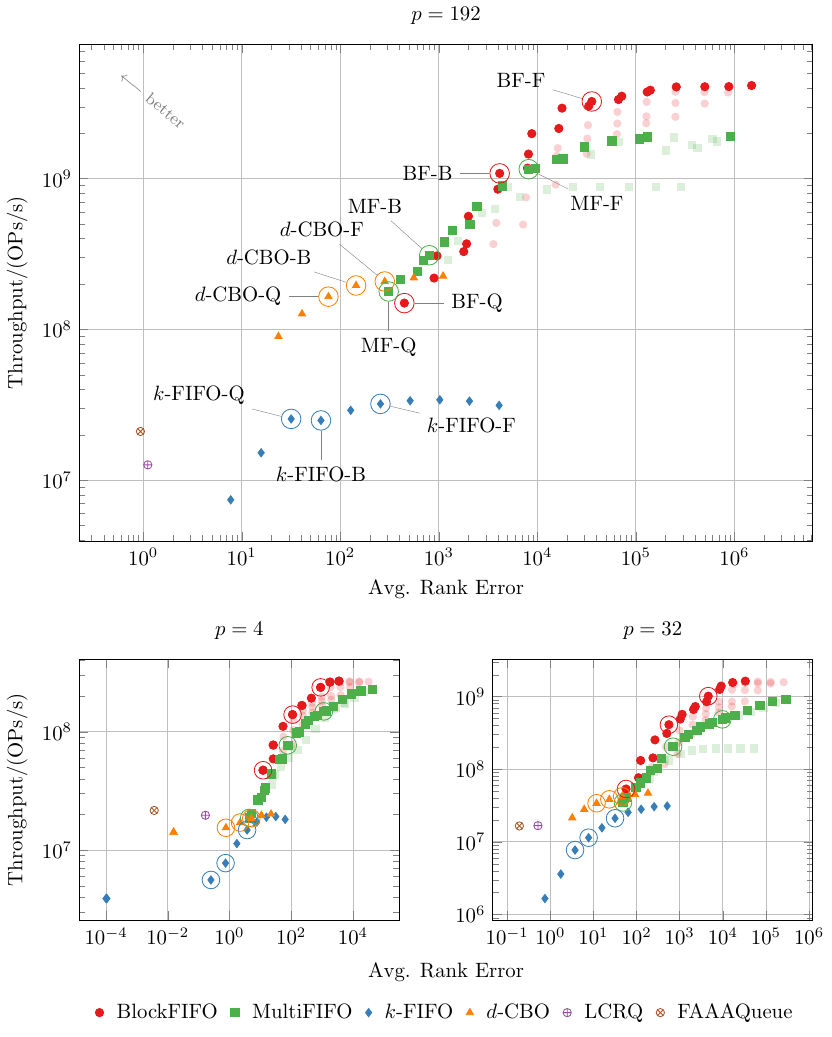}
    \caption{Various configurations of all competitors on the push-pop benchmark with different thread counts on machine \texttt{AMD}.
        For the \blockFIFO{}, the block factor $B$ ranges from \num{1} to \num{16} and the block size $C$ ranges from \num{7} to \num{2047}.
        For the MultiFIFO, sub-queues per thread $c$ range from \num{2} to \num{8} and stickiness $s$ ranges from \num{1} to \num{4096}.
        For the $k$-FIFO, segment size $k$ ranges from $\tfrac{1}{8}p$ to $64p$.
        For the $d$-CBO, the sub-queue count $c$ ranges from $\tfrac{1}{8}p$ to $8p$.
        For all ranges, we sample integer powers of two (minus one for the block size).
        Non-Pareto-optimal configurations of the \blockFIFO{} and MultiFIFO are shown with low opacity.
        Configurations used in further experiments are highlighted with circles and their name is annotated.
    }%
    \label{exp:tuning-all}
\end{figure*}
\Cref{exp:tuning-all} shows the throughput and quality of all competitors for a wide range of parameter configurations on the push-pop benchmark.
The \blockFIFO{} and the MultiFIFO are the only competitors where the throughput increases significantly with higher degrees of relaxation.
Consequently, they achieve an order of magnitude higher throughput than the next fastest competitor (the $d$-CBO) at the cost of higher rank errors.
With lower thread counts ($p=4$, $p=32$), the \blockFIFO{} dominates the MultiFIFO with higher throughput at similar rank errors.
However, with $p=192$ threads, the MultiFIFO achieves higher throughput than the \blockFIFO{} for rank errors below \num{2000}.
With more relaxed configurations, the throughput of the \blockFIFO{} scales faster, outperforming the MultiFIFO significantly.
The $d$-CBO offers slightly lower rank errors than the highest-quality configurations of the MultiFIFO and \blockFIFO{} while achieving similar throughput.
Unfortunately, the throughput of the $d$-CBO barely increases with higher degrees of relaxation.
The $k$-FIFO is dominated by other competitors on the entire Pareto-front.
While it exhibits similar quality to the $d$-CBO, it has significantly lower throughput and also does not scale well with higher degrees of relaxation.
As expected, the strict FIFO queues exhibit virtually no rank errors.
Their throughput is competitive with the highest-quality configurations of the other competitors for low thread counts, but is an order of magnitude slower for \num{192} threads.

For subsequent experiments, we select three configurations (\textbf{Q}uality, \textbf{B}alanced, and \textbf{F}ast) of implementations with tunable quality parameters that are Pareto-optimal on the throughput-quality spectrum for the push-pop benchmark with \num{192} threads on machine \texttt{AMD}.
The selected configurations are given in \cref{tab:comp-config} and additionally annotated in \cref{exp:tuning-all}.
Detailed parameter tuning for the \blockFIFO{} and the MultiFIFO can be found in Appendix~\ref{s:parameter-tuning}.

\begin{table}
    \caption{Selected configurations for configurable competitors used in the experiments.}\label{tab:comp-config}
    \begin{tabular}{c c c c}
        \toprule
        Competitor (Parameters) & Q     & B      & F       \\
        \midrule
        BF        $(B,C)$       & $1,7$ & $1,63$ & $1,511$ \\
        MF        $(c,s)$       & $2,1$ & $4,16$ & $4,256$ \\
        $k$-FIFO  $(k)$         & $p/2$ & $p$    & $4p$    \\
        $d$-CBO $(c)$           & $p/2$ & $p$    & $2p$    \\
        \bottomrule
    \end{tabular}
\end{table}

\begin{figure*}
    \centering
    \includegraphics{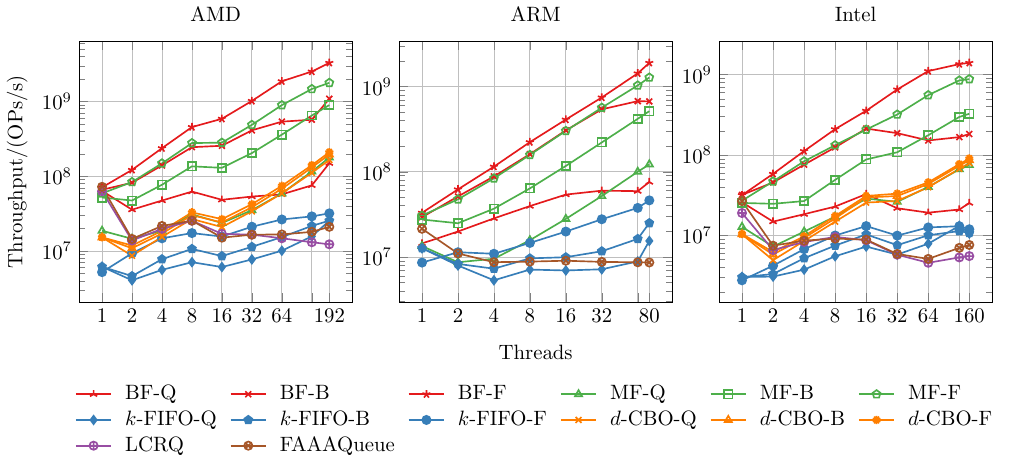}
    \caption{Throughput on the push-pop benchmark with different thread counts on all machines.}%
    \label{exp:performance}
\end{figure*}

\Cref{exp:performance} shows the throughput scaling behavior of all competitors on the push-pop benchmark for all machines.
Unfortunately, the LCRQ and $d$-CBO implementations are not compatible with machine \texttt{ARM}, and are therefore omitted.
The balanced and fast configurations of the \blockFIFO{} and MultiFIFO consistently achieving higher throughput than the other competitors for all thread counts, with the fast \blockFIFO{} being the fastest.
While the $d$-CBO also exhibits some scalability, its fast configuration is an order of magnitude slower than the fast \blockFIFO{} and MultiFIFO configurations.
The quality variants show similar performance to the $d$-CBO with the highest thread count.
The quality configuration of the MultiQueue and the $d$-CBO behave almost identically due to their conceptual similarity.
The $k$-FIFO does not scale well, and---unsurprisingly---the strict queues do not scale at all.
The results are generally consistent across all machines, with the notable exception of the BlockFIFO on the Intel machine.
Here, the quality and balanced configurations do not scale well when using more than one NUMA node ($p>16$).
This is probably due to the fact that all threads must access blocks within the current windows.
These blocks likely reside on the same NUMA node, leading to high bus contention.
The MultiFIFO does not suffer from this problem, since different threads can operate on different sub-queues that may be located on different NUMA nodes.
\Cref{s:quality-scaling} shows the quality scaling behaviour of all competitors on the push-pop benchmark for machine \texttt{AMD}.

\begin{figure*}[t]
    \includegraphics{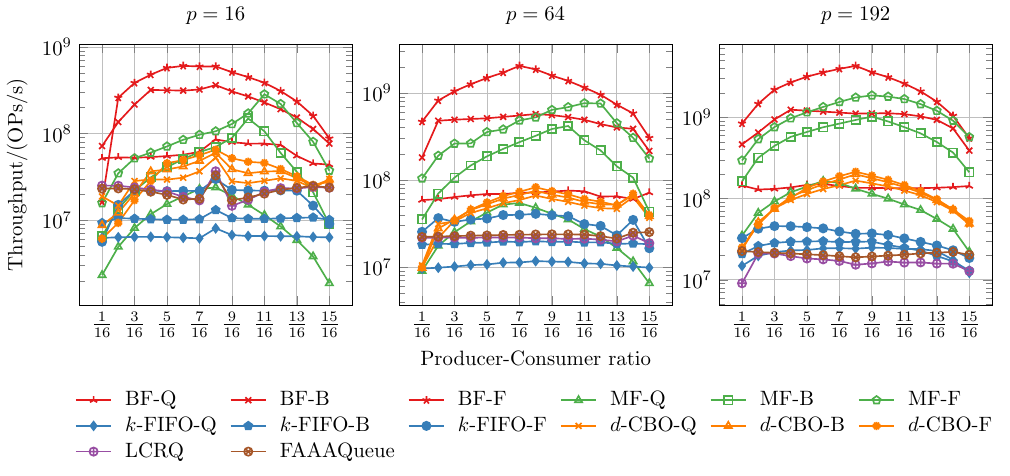}
    \caption{Throughput with different producer-consumer ratios at different thread counts.
    }%
    \label{exp:prodcon}
\end{figure*}

\cref{exp:prodcon} shows the producer-consumer benchmark for different ratios of producers and consumers and different thread counts.
Almost all competitors favor balanced ratios of producers and consumers over extreme imbalances, indicating that the performance of both operations is similar.
Similar to the push-pop benchmark, the balanced and fast variants of the \blockFIFO{} and MultiFIFO generally outperform all other competitors.
Interestingly, the \blockFIFO{} still performs well in consumer-heavy workloads, where the queue is likely to run empty, which is a scenario that it is not designed for.
The MultiFIFO exhibits a bias towards producer-heavy workloads, with an optimal ratio of around $\tfrac{11}{16}$.
However, it has the most pronounced performance drops for extreme ratios among all competitors.
This is likely due to the fact that queues are locked by one type of worker for the majority of the time, while the other worker type starves, as there is no progress guarantee for each type of operation.
The $d$-CBO is more stable across different ratios than the MultiFIFO for all thread counts because it avoids locking by using lock-free sub-queues.
Both perform poorly with consumer-heavy workloads due to triggering a costly emptiness-detection algorithm that scans all sub-queues.
The $k$-FIFO and the strict competitors exhibit very stable performance across all ratios, but they are consistently slower than the other relaxed competitors.

Results of the producer-consumer benchmark on the \texttt{ARM} and \texttt{Intel} machines are given in \cref{s:prodcon_other_machines}.

\subsection{Breadth-First Search.}
The single-source shortest path problem (SSSP) is a fundamental and well-known graph problem.
On unweighted graphs, it can be solved with a Breadth-First Search (BFS).
A straightforward BFS algorithm uses a FIFO queue to store the nodes that are to be explored.
A natural parallelization of this algorithm is then to use a concurrent queue.
Williams and Sanders~\cite{williamsEngineeringMultiQueuesFast2025} describe a parallel SSSP algorithm for weighted graphs that employs a concurrent priority queue.
The idea of the algorithm is to allow the exploration of sub-optimal nodes, potentially requiring the re-exploration of parts of the graph when a shorter path to a node is found later.
We adapt this algorithm to unweighted graphs by using a concurrent FIFO queue instead of a concurrent priority queue.

We evaluate the BFS on various real-world graphs for strong scaling behaviour and on random graphs (generated with KaGen~\cite{funke2017communication}) for weak scaling behaviour.
The real-world graphs are the road networks of Europe and the USA, the follower-relationships on Twitter~\cite{kwak2010twitter}, cross-references in the English Wikipedia and the network of \textit{.uk} domains~\cite{BCSU3}.\footnote{The road network graphs were obtained from \url{https://i11www.iti.kit.edu/resources/roadgraphs.php}, the others from \url{https://law.di.unimi.it/datasets.php}}
\cref{exp:ss_graphs} compares core characteristics of these graphs.
\begin{table}
    \caption{Number of nodes ($n$) and edges ($m$), as well as average node degrees for all graph instances used for strong scaling.}%
    \label{exp:ss_graphs}
    \begin{tabular}{c S[table-format=3.1] S[table-format=4.1] S[table-format=2.1]}
        \toprule
        {Name}                 & {$n/10^6$} & {$m/10^6$} & {Avg\@. deg\@.} \\
        \midrule
        \texttt{enwiki-2022}   & 6.5        & 159.0      & 49.0            \\
        \texttt{ljournal-2008} & 5.4        & 79.0       & 29.5            \\
        \texttt{twitter-2010}  & 41.7       & 1468.4     & 70.5            \\
        \texttt{uk-2005}       & 39.5       & 936.4      & 47.5            \\
        \texttt{OSM Europe}    & 173.8      & 348.0      & 4.0             \\
        \texttt{OSM USA}       & 23.9       & 28.9       & 2.4             \\
        \bottomrule
    \end{tabular}
\end{table}
For the weak scaling experiments, we use the following graph classes:
\begin{itemize}
    \item \texttt{RGG2D}: Random geometric graph with points on the 2D plane.
    \item \texttt{RHG}: Random hyperbolic graph with a gamma of $\gamma=2.7$.
    \item \texttt{GNM}: Erdos-Rényi graph.
\end{itemize}
The number of nodes scales linearly with the number of threads with a scaling factor of $2^{16}$, resulting in $n=2^{16}p$ nodes.
The average vertex degree is set to \num{64}.

\begin{figure*}
    \centering
    \includegraphics{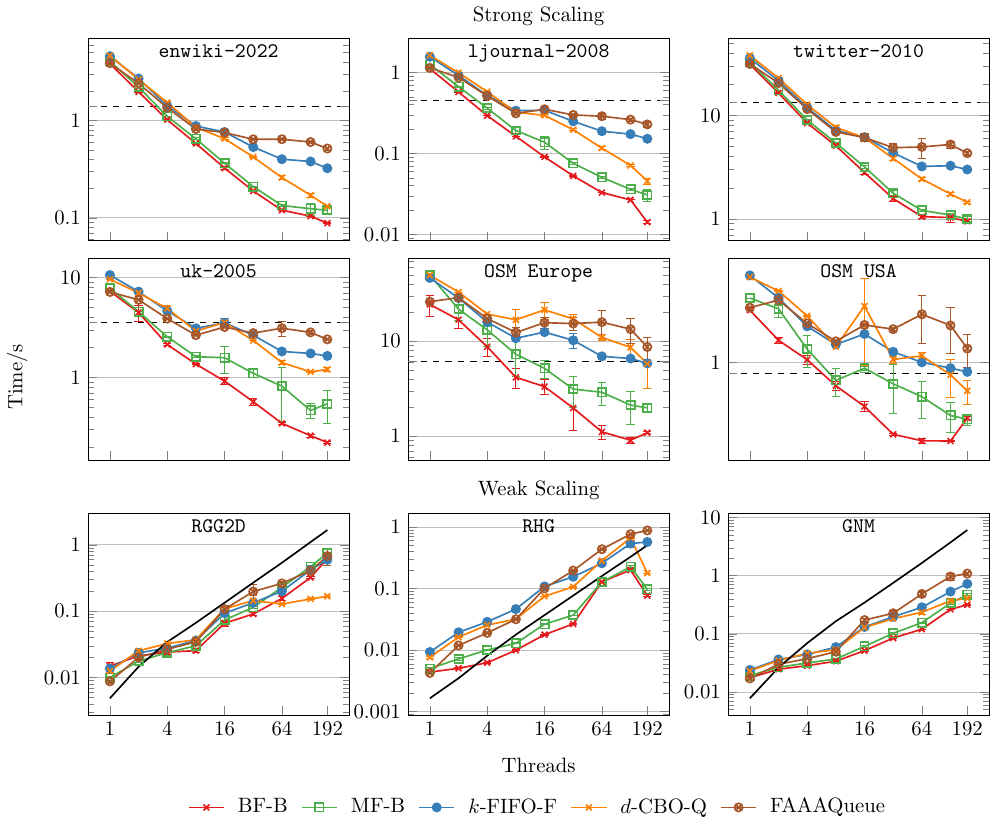}
    \caption{Weak and strong scaling BFS benchmarks.
        Only the best-performing queue configurations for each competitor are shown for clarity.
        The upper part shows strong scaling behaviour on real-world graphs, where the dotted lines represent the execution time of a sequential BFS.
        The bottom part shows weak scaling behaviour, where the graph size scales with the number of threads.
        The black line represents the time required by a sequential BFS on the scaled graph.
    }%
    \label{exp:ss_bfs}
\end{figure*}
\cref{exp:ss_bfs} shows the performance on the strong scaling and weak scaling graphs.
On the real-world graphs, the \blockFIFO{} and MultiFIFO generally scale well and achieve a speedup of up to \num{10} over the sequential BFS, except for the \texttt{OSM USA} graph.
The \blockFIFO{} is consistently the best-performing queue for all thread counts and graph instances.
On the \texttt{enwiki-2022}, \texttt{ljournal-2008} and \texttt{twitter} graphs, the $d$-CBO also has good scalability and competitive performance to the MultiFIFO.
Particularly on the road networks, no competitor besides the \blockFIFO{} and the MultiFIFO is able to substantially outperform the sequential BFS even with 192 threads.
The non-relaxed competitors do not scale beyond \num{8} threads and are consistently slower than all relaxed competitors.

The weak scaling graphs show similar results, except for the \texttt{RGG2D} graph class, where the $d$-CBO outperforms all other competitors for high thread counts.
On all other graphs, the \blockFIFO{} and the MultiFIFO outperform the other competitors consistently.
On the \texttt{RHG} graph, the other competitors are slower than the sequential algorithm except for the $d$-CBO with the highest number of threads.

\begin{figure*}
    \centering
    \includegraphics{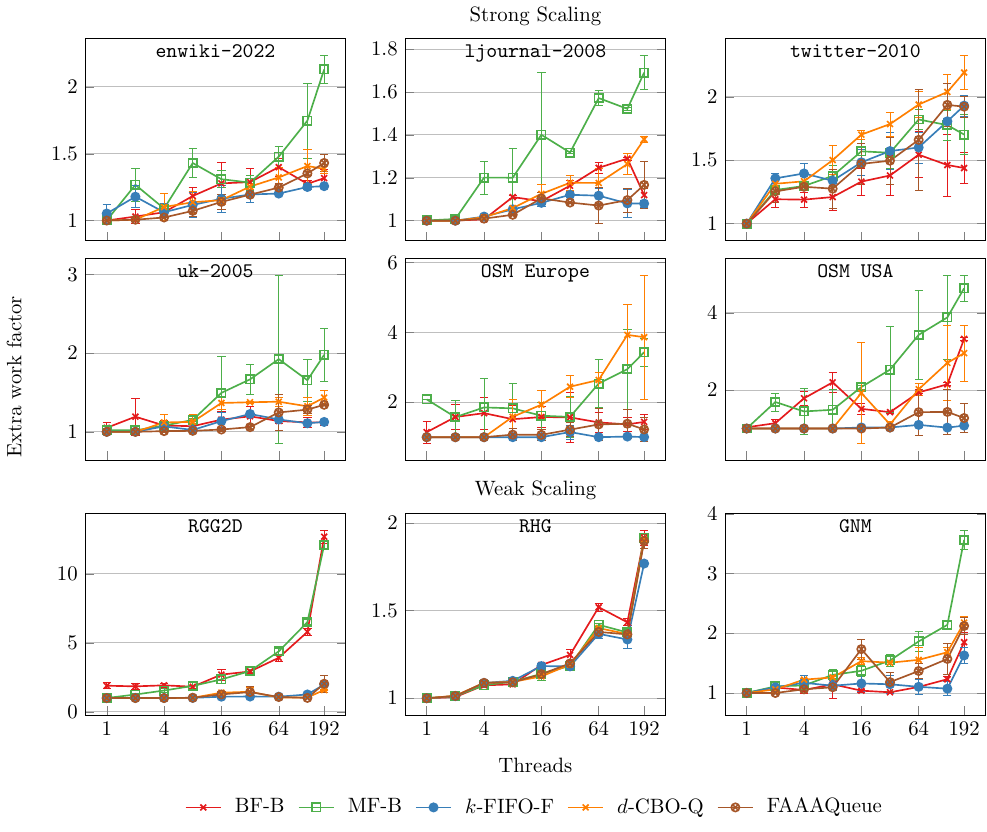}
    \caption{Comparing extra work of weak and strong scaling BFS benchmarks.
        Only the queue configurations from \Cref{exp:ss_bfs} are shown.
    }%
    \label{exp:ss_bfs_work}
\end{figure*}

\cref{exp:ss_bfs_work} shows the extra work incurred by the relaxed queues as the number of processed nodes divided by the number of processed nodes in a sequential BFS.
Generally, the extra work is within one order of magnitude of the sequential BFS.
Since the throughput of the relaxed queues is up to two orders of magnitude higher than strict queues, this is a worthwhile trade-off.
While the MultiFIFO often induces the most extra work, the \blockFIFO{} is often competitive with the strict \texttt{FAAAQueue}.
The \texttt{RGG2D} graph is a notable outlier, where both the \blockFIFO{} and MultiFIFO induce significantly more extra work than the other competitors.
This aligns with the $d$-CBO outperforming them on this graph, but explaining the discrepancy requires further analysis.

\section{Conclusion and Future Work.}
With the MultiFIFO and the BlockFIFO, we have introduced relaxed concurrent FIFOs that allow very high throughput and scale with increased relaxation.
In particular the BlockFIFO still offers many opportunities for further improvement.
Besides giving a ``proper'' theoretical analysis, we would like to (1) make them perform better when almost empty and (2) avoid the quadratic dependence of rank errors on the number of threads.

Problem (1) may be addressed by observing that, with very few elements, we can remove any element and we can insert a new element anywhere while respecting the rank error bounds.
One possibility is to allow overlapping push and pop windows and allow concurrent insertions and deletions in the same block.
However, this approach makes emptiness detection more difficult.

For Problem (2), we consider guiding threads to blocks more efficiently by augmenting the bitsets with a hierarchical structure supporting efficient, low-contention traversal.
This might allow us to transition to blocks whose size need not scale with the number of threads.

We showed that relaxed FIFO queues offer a simple yet promising approach to parallel graph searches.
We intend to investigate their application in further domains, and view a parallel version of the preflow-push algorithm for the maximum-flow problem~\cite{goldbergNewApproachMaximumflow1988} as a promising candidate.
Analyzing the connection between the quality of a queue and the extra work incurred is another interesting direction for future research, which may help to select and tune the appropriate data structure for a given application.

Finally, we find it interesting to look at adaptations of the \blockFIFO{} for other architectures like GPUs or distributed memory.
\section*{Acknowledgments.}
This project has received funding from the European Research Council (ERC) under the European Union’s Horizon 2020 research and innovation program (grant agreement No. \num{882500}).\\
\includegraphics[width=4cm]{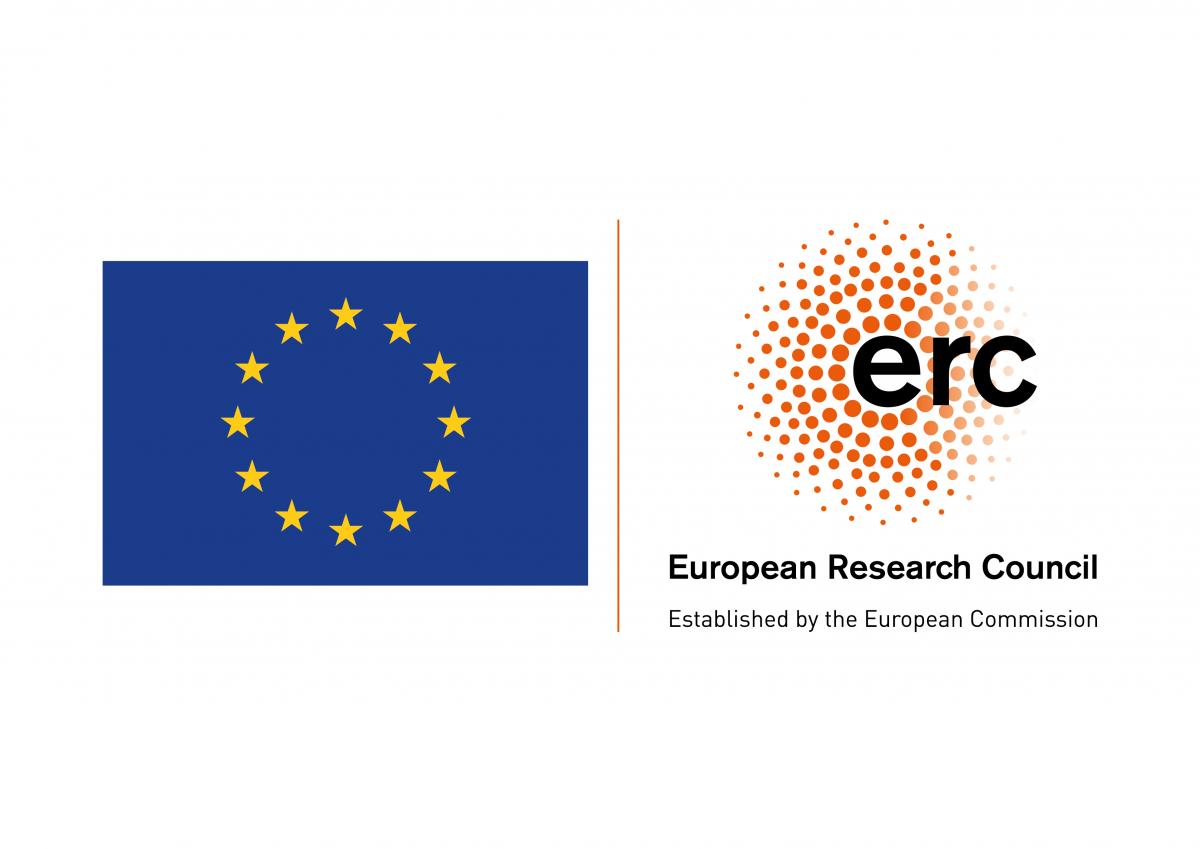}
\bibliography{references}
\clearpage

\appendix
\crefalias{section}{appendix}

\section{Lock-Free Ring Buffer Implementation.}\label{s:bf-impl}

Here, we describe the lock-free implementation of the \blockFIFO{} using a fixed-size ring buffer.
The blocks are stored in a linear array $A$ of size $|A|=k\cdot w$ for some integer $k\geq 3$.
Each block consists of a \emph{header} and an array of $C$ \emph{cells} that store the elements.
The block header is represented as a single integer that encodes an epoch, a pop counter, a push counter, and a bit to indicate whether the block is claimed.
The epoch is a monotonically increasing integer that is incremented whenever the block is closed.
A \emph{block index} is an integer that encodes both the position of a block in the array and its \emph{epoch}.
Specifically, block index $i$ references block $A[i \mod |A|]$ in epoch $\lfloor i / |A|\rfloor$.
The block index $i=-1$ indicates an invalid block by convention.
The push and pop windows are represented as block indices to the first block in the respective window.
A block index is \emph{valid} if it is not $-1$ and the referenced block has the same epoch as the block index.
If the epoch of a block index is smaller than the epoch of the referenced block, the block is regarded as closed.
Each thread maintains block indices to the blocks it last pushed into and popped from, respectively.
Pseudocode for the \Op{push} and \Op{pop} operations is provided in Algorithm~\ref{algo:ins} and~\ref{algo:del}, respectively.
\begin{algorithm2e*}
    \caption{Pseudocode for the \Op{push} operation.}\label{algo:ins}
    \SetKwFunction{Push}{push}
    \SetKwFunction{InsertInBlock}{insertInBlock}
    \SetKwFunction{ClaimBlock}{claimBlock}
    \SetKwData{PushBlock}{pushBlock}
    \KwResult{\True if the insertion succeeds; \False if the queue is full}
    \KwData{\PushBlock, the block index of the last block the thread inserted into}
    \KwIn{Element to insert $e$}
    \BlankLine{}
    \Fn{\Push{$e$}}{
        $p \gets \Load{\PushWindow}$\;
        \If{$\PushBlock \geq p$}{
            $h \gets \Load{\Array{\GetPos{\PushBlock}}.\Header}$\;
            \If{$h.\Epoch = \GetEpoch{\PushBlock} \land \InsertInBlock{$h$, \PushBlock, $e$}$}{
                \lIf{$h.\PushCounter = C - 1$}{\PushBlock{} $\gets -1$}
                \KwRet{\True}\;
            }
        }

        \RepeatInf{
            $r \gets \Random{$0$, $w-1$}$\;
            \For(\Comment*[f]{search random unclaimed block}){$j \gets 0$ \KwTo{} $w-1$}{
                $i \gets p + (r + j \mod w)$\;
                $h \gets \Load{\Array{\GetPos{$i$}}.\Header}$\;
                $h^* \gets \ToHeader{\GetEpoch{$i$}, $0$, $0$, \False}$\;
                $h' \gets \ToHeader{\GetEpoch{$i$}, $0$, $0$, \True}$\;
                \If(\Comment*[f]{block claimed}){$h = h^* \land \CAS{\Array{\GetPos{$i$}}.\Header, $h$, $h'$}$} {
                    \lIf{\InsertInBlock{$h$, $i$, $e$}}{
                        $\PushBlock \gets i$; \KwRet{\True}
                    }
                }
            }
            \If(\Comment*[f]{queue is full}){$p + w - \Load{\PopWindow} = \left|\Array\right|$}{
                $\PushBlock \gets -1$; \KwRet{\False}
            }
            $p \gets \CAS{$\PushWindow$, $p$, $p + w$}$\Comment*[r]{advance push window}
        }
    }
    \BlankLine{}
    \Fn{\InsertInBlock{$h$, $i$, $e$}}{
        \If{\CAS{\Array{\GetPos{$i$}}.\Cells{$h.\PushCounter$}, \Empty, $e$}}{
            $h' \gets \ToHeader{$h.\Epoch$, $0$, $h.\PushCounter + 1$, \True}$\;
            \lIf{\CAS{\Array{\GetPos{$i$}}.\Header, $h$, $h'$}}{\KwRet{\True}}
            $\Array{\GetPos{$i$}}.\Cells{$h.\PushCounter$} \gets \Empty$\;
        }
        \KwRet{\False}\;
    }
\end{algorithm2e*}

\begin{algorithm2e*}
    \caption{Pseudocode for the \Op{pop} operation.}\label{algo:del}
    \SetKwFunction{Pop}{pop}
    \SetKwFunction{FindOpenBlock}{findOpenBlock}
    \SetKwFunction{ReserveElement}{reserveElement}
    \SetKwData{PopBlock}{popBlock}
    \SetKwData{OpenBlock}{openBlock}
    \SetKwData{PushWindowEmpty}{pushWindowEmpty}
    \KwResult{The deleted element $e$, or $\bot$ if the queue is empty.}
    \KwData{\PopBlock, the block index of the last block the thread deleted from}
    \BlankLine{}
    \Fn{\Pop{}}{
        \If{$\PopBlock \neq -1$}{
            $h \gets \Load{\Array{\GetPos{\PopBlock}}.\Header}$\;
            \If{$h.\Epoch = \GetEpoch{\PopBlock} \land \ReserveElement{$h$, \PopBlock}$}{
                \KwRet{\Swap{\Array{\GetPos{\PopBlock}}.\Cells{$h.\PopCounter$}, \Empty}}\;
            }
        }
        \RepeatInf{
            $p \gets \Load{\PopWindow}$; $q \gets \Load{\PushWindow}$\;
            \If{$p + w < q \land \Load{\Array{\GetPos{$p$}}.\Header}.\Epoch \neq \GetEpoch{$p$}$}{
                \CAS{$\PopWindow$, $p$, $p+1$}; \Continue\Comment*[r]{first block is closed}
            }
            $r \gets \Random{$0$, $w-1$}$\;
            \For{$j \gets 0$ \KwTo{} $w-1$}{
                $i \gets p + (r + j \mod w)$\;
                \RepeatInf{
                    $h \gets \Load{\Array{\GetPos{$i$}}.\Header}$\;
                    \lIf{$h.\Epoch \neq \GetEpoch{$i$}$}{\Break}
                    \If{$\ReserveElement{$h$, \PopBlock} \land h.\PushCounter > 0$}{
                        $\PopBlock \gets i$; \KwRet{\Swap{\Array{\GetPos{$i$}}.\Cells{$h.\PopCounter$}, \Empty}}\;
                    }
                }
            }
            \If(\Comment*[f]{pop window is directly behind push window}){$p + w = q$}{
                $\PushWindowEmpty \gets \True$\;
                \For(\Comment*[f]{scan push window for elements}){$i \gets q$ \KwTo{} $q+w-1$}{
                    $h \gets \Load{\Array{\GetPos{$i$}}.\Header}$\;
                    \lIf{$h.\PushCounter > 0$}{ $\PushWindowEmpty \gets \False$; \Break{}}
                }
                \lIf{$\PushWindowEmpty \land q = \Load{\PushWindow}$}{\KwRet{\Empty}}
                \CAS{$\PushWindow$, $q$, $q + w$}\Comment*[r]{advance push window}
                \CAS{$\PopWindow$, $p$, $p + w$}\Comment*[r]{advance pop window}
            }
        }
    }
    \BlankLine{}
    \Fn{\ReserveElement{$h$, $i$}}{
        \uIf(\Comment*[f]{block has more than one elements}){$h.\PopCounter + 1 < h.\PushCounter$}{
            $h' \gets \ToHeader{$h.\Epoch$, $h.\PopCounter + 1$, $h.\PushCounter$, \True}$\;
        }\Else(\Comment*[f]{block can be closed}){
            $h' \gets \ToHeader{$h.\Epoch + 1$, $0$, $0$, \False}$
        }
        \KwRet{\CAS{\Array{\GetPos{$i$}}.\Header, $h$, $h'$}}\;
    }
\end{algorithm2e*}

To claim a block, a thread changes the block's claim bit from \DataSty{false} to \DataSty{true}.
Inserting an element into a claimed block involves three steps.
First, the threads reads the block header and checks that the block index is valid and that the block is not full.
Second, it attempts to place the element into the cell indicated by the push counter, checking that the cell was empty (i.e., contained $\bot$) prior to the operation.
Third, the thread attempts to \emph{commit} the element by incrementing the push counter in the header if the header remained unchanged during the operation.
If committing fails, the operation is reverted by resetting the cell to $\bot$.

If any of the three steps fails, the thread attempts to claim another random unclaimed block in the push window.
If the push window contains no unclaimed blocks, the thread advances the push window by $w$ blocks, or, if the queue is full, the insertion fails.

When deleting an element from a non-empty block, a thread first attempts to \emph{reserve} the element.
If it was the last element in the block, it does so by incrementing the epoch counter, thereby closing the block.
Otherwise, it does so by incrementing the block's pop counter.
After reserving an element, the thread atomically retrieves the element and replaces its cell with $\bot$.

Crucially, an inserted element cannot be deleted until it is committed:
Deletions only consider cells with indices within the block's push and pop counters.
Once the element is inserted into its cell, no other thread can advance the block's push counter beyond the cell's index, since subsequent insertions into that cell would fail.
Moreover, if the push counter was already greater than the cell's index, then the pop counter must also have been greater than the cell's index; otherwise, the cell would not have been empty when the element was inserted.
If a thread suspends after writing an element to a cell but before committing it, only the cells up to this cell can be used in later epochs.
When the thread resumes, it will either commit the element or revert the operation, making all cells in the block available again.
As soon as a deleting thread reserves an element in a block, no other thread can reserve the same element.
If a deleting thread suspends right after reserving an element but before deleting it, the remaining elements in the block can still be deleted and the block can still be closed.
Again, only the cells up to the reserved element can be used in later epochs.
When the thread resumes, it will delete the element, making all cells in the block available again.

\subsection{Lock-Freedom.}

We sketch the main argument for lock-freedom, showing that within a bounded number of steps, at least one thread is guaranteed to complete its operation.

An inserting thread will, within a bounded number of steps, either complete its operation, attempt to insert an element into a claimed block, attempt to claim a new block, or attempt to advance the push window.
When inserting an element succeeds, the operation completes in a bounded number of steps.
If inserting an element into a claimed block fails, it attempts to claim a new block.
Since each block can be claimed at most once per epoch, one inserting thread will, within a bounded number of steps, either complete the insertion (potentially failing) or successfully advance the push window.
Once the push window is advanced, all blocks in the new push window are unclaimed.
The pop window cannot advance to these blocks before at least one insertion into a block in the new push window is completed.
Since insertions never update the header or place elements in the cells of a block claimed by other threads, at least one thread must be able to complete its operation within a bounded number of steps.

A deleting thread will, within a bounded number of steps, either complete its operation, attempt to reserve an element to delete, attempt to close a block, or attempt to advance the pop window.
When reserving an element succeeds, the operation completes in a bounded number of steps.
If reserving an element or closing a block fails, it is retried until it succeeds or the block is closed.
However, the block header can be changed at most $C+1$ times by inserting threads before a deletion successfully reserves an element or closes the block.
Since each block can be closed at most once per epoch, one deleting thread will, within a bounded number of steps, either complete its operation (potentially failing) or successfully advance the pop window.
Thus, if at least one element is in the queue, after a bounded number pop window advances, at least one block in the pop window will contain an element.
After a bounded number of steps, at least one deleting thread will be able to reserve an element in the pop window and complete its operation.

\section{Bitset Details.}\label{s:bitset-details}

The bitset is partitioned into small atomic units.
In order to prevent false sharing, atomic units are aligned to cache lines.
It is desirable to establish the bitset as an arbiter of truth to actually allow for avoiding potentially numerous block header reads.
This necessitates that the bitset must only exhibit one-sided errors, where a bit may be set even if the corresponding block does not contain any elements.
Some kind of one-sided error is unavoidable, as block push/pop operations must be ordered in some way with the corresponding bitset modifications.
By choosing the side of the errors like we have, it enables deleters to concern themselves only with the set bits, which they eventually reset over a bounded number of delete operations.
Blocks associated with unset bits are of no concern to them.
Additionally, it is necessary to bundle an epoch with the bitset, which can be stored and modified alongside each atomic unit.
Without epochs, deleters may create a two-sided error by resetting the bit of a block that contained no elements when they have started the operation, but has since been filled in the subsequent epoch.

The search operation on each atomic unit can be efficiently implemented using a fixed amount of instructions independently of the size of the atomic unit $u$.
This implementation relies on the bitwise rotation and count leading zeroes instructions commonly available on CISC architectures.\footnote{For example \Op{ROR} and \Op{LZCNT} on \texttt{x86}.}
The search operation over the entire bitset simply encompasses a linear iteration over all atomic units, executing the atomic unit search operation on each until a desired bit is found.

This means that $u$ is a tuning parameter, offering a trade-off between reducing the contention on the individual atomic units with a small $u$ and allowing more blocks to be checked in the same amount of operations with large $u$.
In practice, choosing $u$ to be minimal has shown to provide the best performance until very high degrees of relaxation, where contention becomes so low that the reduced operation count proves more beneficial.

\section{Parameter Tuning.}\label{s:parameter-tuning}
We investigate the trade-off between quality and performance offered by different queues and their configurations and select specific configurations for further experiments.
For this purpose, we use the push-pop benchmark with the maximum of \num{192} threads on machine \texttt{AMD} to measure both throughput and quality.

\paragraph*{MultiFIFO.}

\begin{figure}
    \centering
    \includegraphics{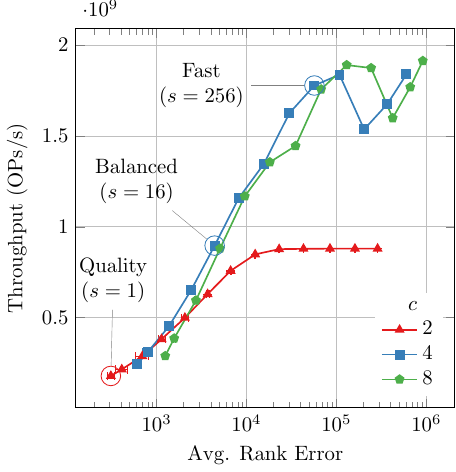}
    \caption{MultiFIFO parameter tuning of the sub-queues per thread $c$ and stickiness period $s$ using the push-pop benchmark.
        Stickiness ranges from 1 to 4096.
        Annotated data points are configurations used in further experiments.%
    }\label{exp:multififo_tune}
\end{figure}

In \cref{exp:multififo_tune}, we compare different configurations of the MultiFIFO.
We vary the sub-queues per thread $c$ between 2, 4 and 8 and the stickiness period between 1 (which is equivalent to no sticking behavior) and 4096, only considering values that are a power of two.

The MultiFIFO plateaus relatively quickly with increasing stickiness, offering no better performance at continuously decreasing quality.
Doubling the queues per thread from 2 to 4 doubles the achievable performance, however, further doubling the queues per thread from 4 to 8 has no beneficial effect on either performance or quality.

\paragraph*{\blockFIFO{}.}

\begin{figure}
    \centering
    \includegraphics{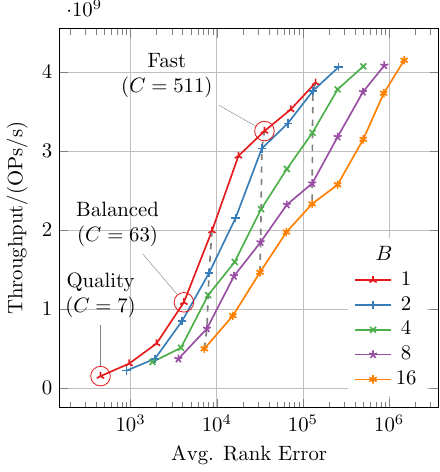}
    \caption{\blockFIFO{} parameter tuning of the blocks per window per thread $B$ and cells per block $C$ using the push-pop benchmark.
        Annotated data points are configurations used in further experiments.
        The gray dashed lines indicate configurations with the same total number of cells in the windows (i.e., the product of $B$ and $C$ is the same).%
    }\label{exp:bbq_tune}
\end{figure}

In \cref{exp:bbq_tune} we compare different configurations of the \blockFIFO{}.
We tune the parameters $B$ and $C$.
Only powers of two are tested, with $B$ ranging from 1 to 16 and $C$ ranging from 7 to 2047.
Notably, $C$ is always $2^x-1$ in order to accommodate for the 64-bit block header.

The \blockFIFO{} exhibits a peculiar vertical clustering of points around one level of quality as illustrated by the dashed lines.
This behavior can be sufficiently explained by quality correlating to the amount of \textit{cells} per window, so the product of window size $w$ and block size $C$.\footnote{There are always $w$ less cells per window due to the block headers, but that is irrelevant in practice.}
For a fixed amount of cells per \textit{window} a low $B$ and high $C$ appears optimal.
Such a configuration minimizes the amount of potentially high-contention block claim operations necessary to fill/empty an entire window, while maximizing the time that can be spent in cache-friendly, typically low-contention inner-block operations.
Additionally, a low $B$ minimizes the maximum possible time spent looking for a valid block within a window by minimizing the amount of blocks there are.

\section{Multi-Start BFS.}\label{sec:msbfs}
On graph instances where scalability is limited, more significant speedups can be achieved by executing multiple searches concurrently within a single queue.
We start searches from $s$ nodes, storing $s$ distinct distance arrays and including the associated source node index in the elements pushed to the queues.
This is reminiscent to the pre-computation required for an $A$* with landmarks (ALT) search~\cite{goldberg2005computing} without weights.

\begin{figure}
    \centering
    \includegraphics{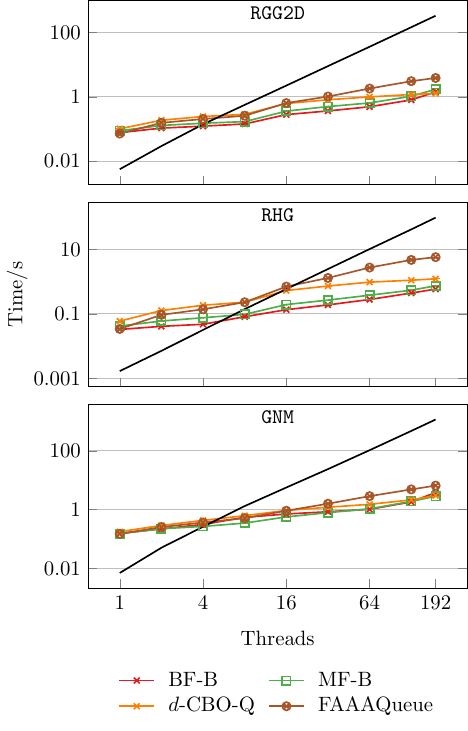}
    \caption{Weak scaling BFS benchmarks with $s=8$ concurrent searches per instance.
        Graph instances are the same as used in \cref{exp:ss_bfs}.
        The $k$-FIFO is unable to complete this benchmark due to its inability to store a full 64 bits of data in its elements which is required to account for the source node indices.
    }\label{exp:multistart}
\end{figure}

As shown in \cref{exp:multistart}, the scalability is vastly superior to a single search being executed.
Especially the MultiFIFO and \blockFIFO{} benefit from the additional work, achieving a speedup of two orders of magnitude with the highest number of threads.
While the $d$-CBO outperformed the \blockFIFO{} and MultiFIFO on the \texttt{RGG2D} graph on the standard BFS benchmark, this is not the case for the multi-start BFS.

\section{Complete Benchmarks.}\label{sec:complete-benchmarks}
\paragraph*{Quality.}\label{s:quality-scaling}
\begin{figure}
    \includegraphics{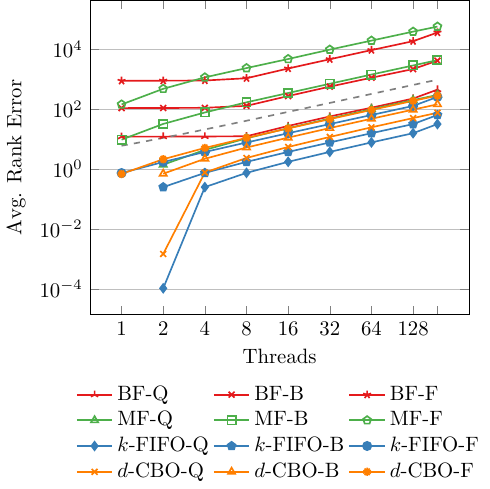}
    \caption{The rank errors during the push-pop benchmark at different thread counts.
        Some configurations exhibit no rank errors with a single thread, these data points are omitted.
        For reference, we include the linear function $5p+1$ as a dashed line.
    }%
    \label{exp:quality}
\end{figure}

\cref{exp:quality} shows the rank error behavior of the relaxed competitors on the push-pop benchmark for different thread counts.
All competitors demonstrate the expected linear scaling of rank errors for all competitors.

\paragraph*{Producer-Consumer on other Machines.}\label{s:prodcon_other_machines}
\begin{figure*}
    \includegraphics{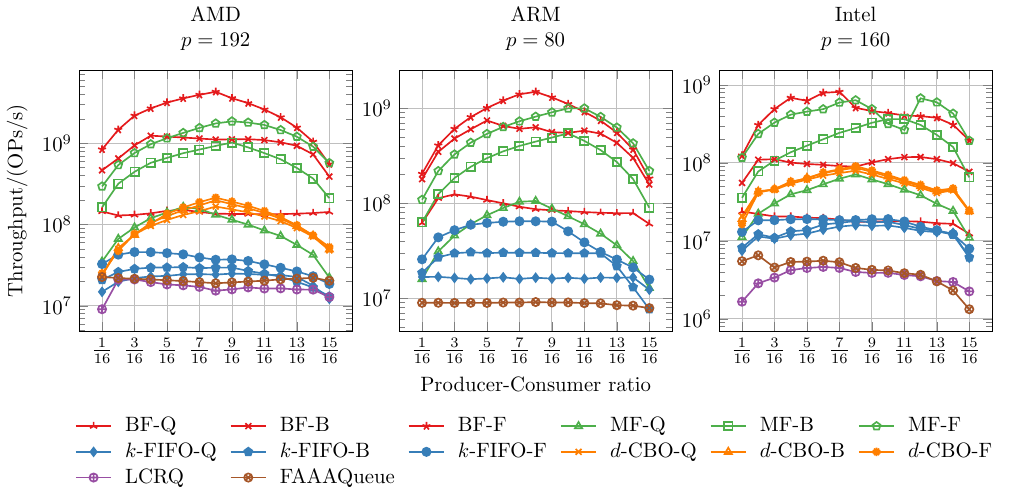}
    \caption{Measuring throughput with different producer-consumer ratios on other machines.}%
    \label{exp:prodcon_other_machines}
\end{figure*}
The behavior is largely similar on the other machines, as seen in \cref{exp:prodcon_other_machines}.
The \blockFIFO{} performance degrades on the Intel machine, with a stark drop in performance for producer-heavy workloads for the fast \blockFIFO{} that is also similarly displayed by the fast MultiFIFO.

\paragraph*{BFS with all competitors.}\label{s:bfs_all_queues}
\begin{figure*}
    \includegraphics{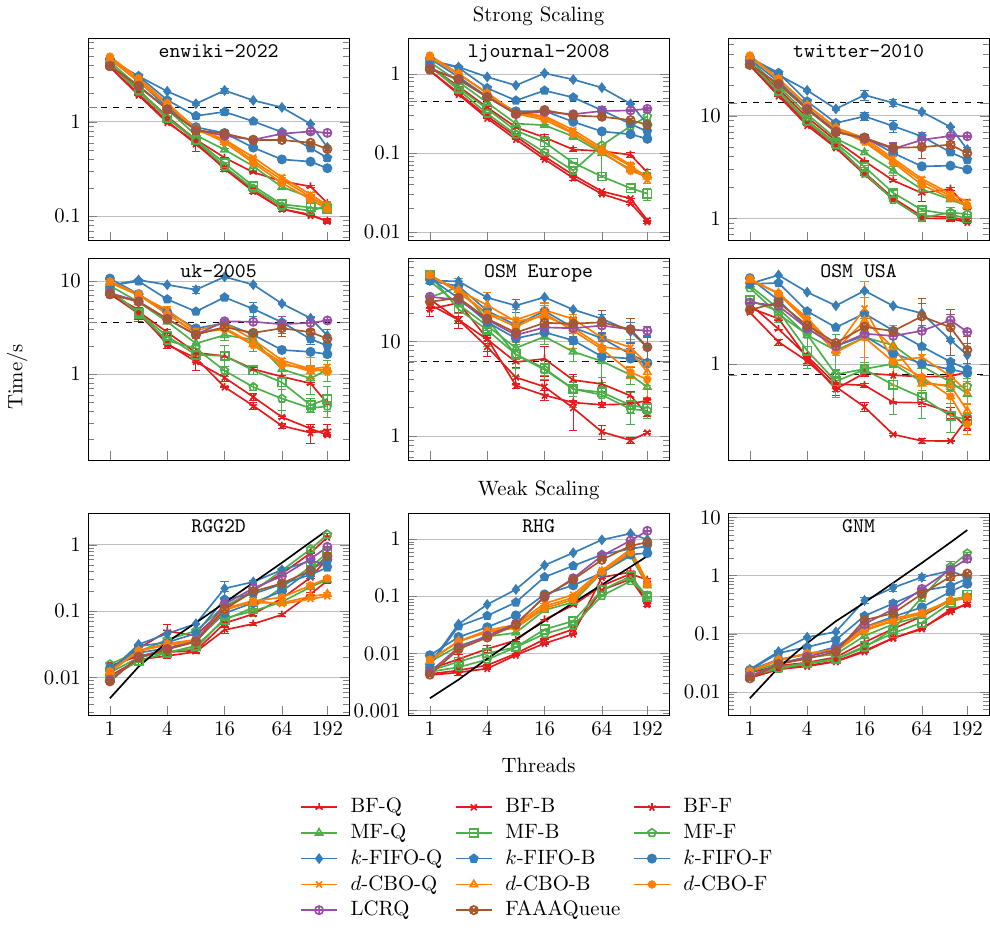}
    \caption{Equivalent to \Cref{exp:ss_bfs} with all queues and configurations shown.}
    \label{exp:ss_bfs_all}
\end{figure*}
\begin{figure*}
    \includegraphics{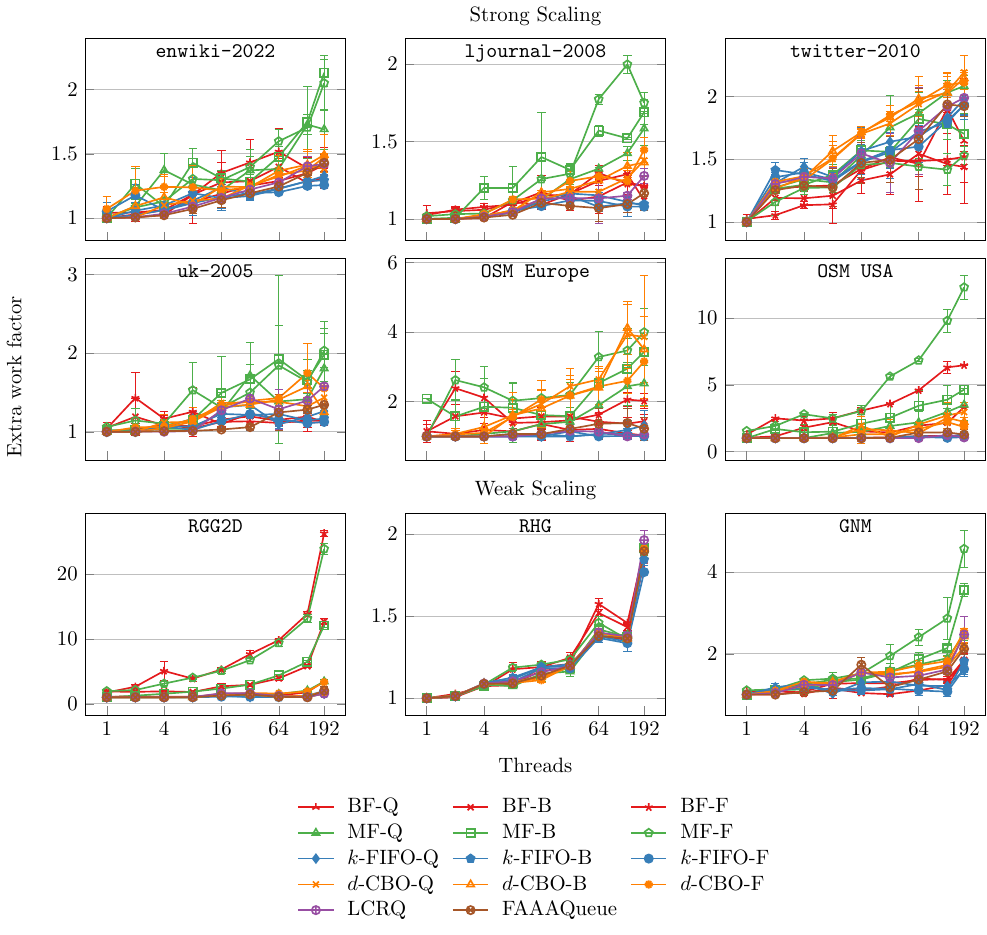}
    \caption{Equivalent to \Cref{exp:ss_bfs_work} with all queues and configurations shown.}
    \label{exp:ss_bfs_work_all}
\end{figure*}
\cref{exp:ss_bfs_all} and \cref{exp:ss_bfs_work_all} show the BFS benchmarks with all competitors and configurations.
\end{document}